\documentclass[a4paper,UKenglish]{lipics-v2018}

\usepackage{microtype}
\usepackage[linesnumbered,boxed]{algorithm2e}

\title{A note on self-improving sorting with hidden partitions}

\author{Siu-Wing Cheng$^\dag$, Man-Kwun Chiu$^\ddag$, Kai Jin$^\dag$}{$^\dag$ HKUST, Hong Kong. \qquad $^\ddag$ Freie University\``{a}t Berlin, Germany}{}{}{}

\authorrunning{S. Cheung, K. Jin}
\Copyright{Siu-Wing cheung and and Kai Jin}
\subjclass{Theory of computation}
\keywords{Self-improving algorithm}

\EventEditors{}
\EventNoEds{1}
\EventLongTitle{Asian Association for Algorithms and Computation 2019}
\EventShortTitle{AAAC 2019}
\EventAcronym{AAAC19}
\EventYear{2019}
\EventDate{April 19-21, 2019}
\EventLocation{Seoul, South Korea}
\EventLogo{}
\SeriesVolume{}
\ArticleNo{1}
\nolinenumbers
\hideLIPIcs

\begin{document}

\maketitle

\section{Introduction.} The sorting problem under a so-called ``self-improving computational model'' was studied in \cite{best}:
In this model, we will have input instances $I_1,I_2,\ldots,$ etc generated as follows. An instance $I$ contains $n$ elements $x_1^{I},\ldots,x_n^{I}$,
  and its $i$-th ($1\leq i\leq n$) element $x_i^I$ is generated according to a distribution $\mathcal{D}_i$.
    The $n$ distributions $\mathcal{D}_1,\ldots,\mathcal{D}_n$ are fixed but are not given.
    The target is to compute and output $\pi(I)$ -- the ranks of the $n$ elements in $I$.

Let $H(\pi(I))$ denote the entropy of the output $\pi(I)$. The authors in \cite{best} showed that they can design a \emph{learning phase} which learns the distributions and builds some data structures by analyzing several instances so that for a given $I$ in the \emph{operation phase}, they can compute $\pi(I)$ in $O(H(\pi(I))+n)$ expected time, which matches the information theory lower bound.

\smallskip We study in this paper a more general setting which allows some dependency among the $n$ elements.
We assume that the $n$ elements are partitioned into $g$ groups (each element belongs to exactly one group) and
   in the $k$-th ($1\leq k\leq g$) group there is a variable $z_k$ which is generated according to a fixed distribution $\mathcal{D}_k$ and each element in this group is a function of $z_k$. Note that the partition as well as the $g$ distributions $\mathcal{D}_1,\ldots,\mathcal{D}_g$ are not given.

However, we need to impose some constraints on these functions of $z_k$. Assume that the $k$-th group contains $n_k$ elements $x_1,\ldots,x_{n_k}$ and moreover $x_1=f_1(z_k),\ldots,x_{n_k}=f_{n_k}(z_k)$.
    We assume that each function $f_i()$ can have at most $\mu$ extremal points and every pair of functions $f_i()$ and $f_j()$ can have at most $\sigma$ intersections, where $\mu$ and $\sigma$ are known constants.

\smallskip Under such constraints, our result is the following.

\begin{theorem}\label{thm:main}
In operation phase, we can compute $\pi(I)$ in $O(H(\pi(I))+n)$ expected time.
\end{theorem}

\subsection{Technique overview}

\newcommand{\po}{\mathsf{po}}

\subparagraph{Learning phase overview.} We learn the hidden partition using constant many instances.
Also, we construct the $V$-list in the same way as in \cite{best}.
Precisely, take $\lambda=\lceil \log n \rceil$ instances and merge all the $\lambda\cdot n$ elements in these instances into a big list and sort them in increasing order; denote the results by $y_1,\ldots,y_{\lambda n}$. Assign $V_r=y_{r\cdot \lambda} (1\leq r\leq n)$, $V_0=-\infty$, and $V_{n+1}=+\infty$.
We call $V_r$ the \emph{predecessor} of $x_i$ if $x_i\in [V_r,V_{r+1})$. For the $k$-th $(1\leq k\leq g)$ group,
  the predecessors of the $n_k$ elements in this group respectively and the order between these elements are denote by $\po_k$;
  its entropy denoted by $H(\po_k)$.
  Finally, let $n'=\max_kn_k$, and we sample $T=n'(n(\mu+1)+n'\sigma)\log n$ instances to learn the distribution of $\po_k$.

\subparagraph{Operation phase.}
First, we compute $\po_k$ for each $k~(1\leq k\leq g)$.
Second, for each $k$, denote $\sigma_k$ the list of $n_k$ elements in $k$-th group in sorted order,
  find all $r$ such that $\sigma_k\cap [V_r,V_{r+1})$ is nonempty,
    and put the sublist $\sigma_k\cap [V_r,V_{r+1})$ into $S_r$ (So $S_r$ is a set of sublists).
Third, we use a merge sort to merge all the sublists in $S_r$ into one list $s_r$ in sorted order.
Finally, by concatenating $s_0,\ldots,s_n$, we obtain the sorted list of all elements.

\subsection{Running time analysis of the operation phase.}

We need the following three crucial lemmas.

\begin{lemma}\label{lemma:H_k}
For each $k~(1\leq k\leq g)$, we can compute $\po_k$ in $O(H(\po_k)+n_k)$ time.
\end{lemma}

\begin{lemma}\label{lemma:H_sum}
$\sum_k H(\po_k) = H(\pi(I)) + O(n)$.
\end{lemma}

\begin{lemma}\label{lemma:S_r}
With high probability, on our construction of the $V$-list, it is guaranteed that
  for each $r$, the expected size of $S_r$ (i.e. the number of sublists in $S_r$) is a constant.
\end{lemma}

By Lemma~\ref{lemma:H_k}, the first step runs in $O\left(\sum_k H(\po_k)+n_k\right)$ time,
  which is $O(\sum_k H(\po_k))+O(n)=H(\pi(I)) + O(n)$ time further according to Lemma~\ref{lemma:H_sum}.
  The second and last step cost $O(n)$ time.
  The third step takes $O(n)$ time by applying Lemma~\ref{lemma:S_r}. Thus we get Theorem~\ref{thm:main}.

\smallskip Lemma~\ref{lemma:H_sum} follows from Lemma~2.3 of \cite{best} because we can compute $(\po_1,\ldots,\po_g)$ in $O(n)$ comparisons given $\pi(I)$.
Lemma~\ref{lemma:S_r} is the same as Lemma~6 in \cite{isaac18ext}. Lemma~\ref{lemma:H_k} is proved below.

\section{Learning phase I -- compute the hidden partition in $\mu^4$ rounds}

Assume we want to determine whether ($x_1$, $x_2$) is in the same group.

Recall that each function has at most $\mu$ extremal points.
We take $m=\mu^4$ samples of $(x_1,x_2)$. Denote the values by $(x_{1,1},x_{2,1}),\ldots,(x_{1,m},x_{2,m})$.
Without loss of generality, assume that $x_{1,1}\leq x_{1,2}\leq \ldots\leq x_{1,m}$. (Otherwise we make it so by sorting)

Moreover, for any sequence of numbers $(A_1,\ldots,A_m)$ with length $m$,
  we define function $D(A_1,\ldots,A_m)$ as the minimum number $d$ such that
  $(A_1,\ldots,A_m)$ can be partitioned into $d$ monotonic sub-sequence.
A sub-sequence is monotonic if it is either increasing or decreasing.

We can prove that
\begin{itemize}
\item If $x_1$ and $x_2$ are in the same group, $D(x_{2,1},\ldots,x_{2,m})\leq 2\mu+1$;
\item If $x_1$ and $x_2$ are in different groups, $D(x_{2,1},\ldots,x_{2,m})=\Omega(\mu^2)$.
\end{itemize}
Therefore,
\begin{itemize}
\item If $D(x_{2,1},\ldots,x_{2,m})\leq 2\mu+1$, with high probability $(x_1,x_2)$ are in the same group.
\item If $D(x_{2,1},\ldots,x_{2,m})>2\mu+1$, it is definitely true that $(x_1,x_2)$ are in different groups.
\end{itemize}
As a consequence, we can learn the hidden partition easily by calling function $D$.

Moreover, since $\mu$ is a constant, so as $m$, hence it only costs constant time to compute $D$.

\section{Learning phase II -- learn the distribution of $\po_k$}\label{sect:learnII}

We need to introduce some notation here.

For convenience, assume that $x_1,\ldots,x_{n_k}$ are in the $k$-th group.

\begin{figure}[h]
  \centering \includegraphics[width=.6\textwidth]{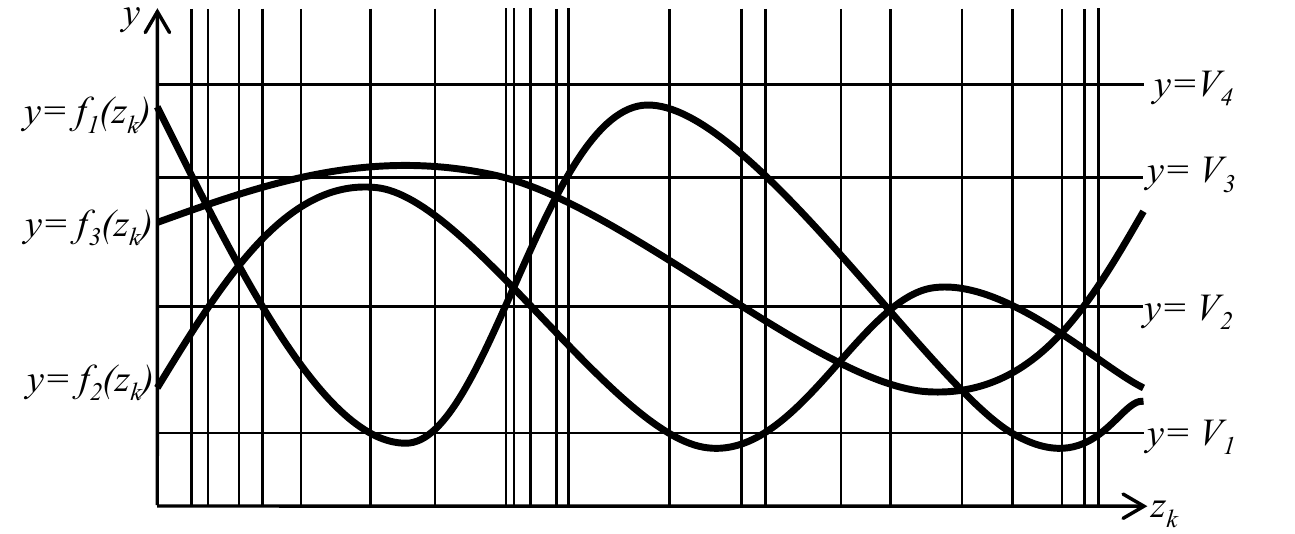}\\
  \caption{Illustration of the arrangement.}\label{fig:arrangement}
\end{figure}

First, we draw $n_k$ curves $y=f_1(z),\ldots,y=f_{n_k}(z)$.
Moreover, for each $r~(1\leq r\leq n)$, we draw a horizontal line $y=V_r$.
Let $\mathcal{A}$ denote the arrangement of these $n+n_k$ curves.

For each intersection in $\mathcal{A}$, we draw a vertical line, as shown in Figure~\ref{fig:arrangement}.
According to our assumption on the functions, there are less than $W=n_kn(\mu+1)+n_k^2\sigma$ such intersections.
These intersections divide the plane into at most $W$ \emph{slabs}.
Notice that $\po_k$ remains the same when $z_k$ is restricted to any fixed slab, yet it could be the same for different slabs.
Thus there are at most $W$ possible (different) choices of $\po_k$, denoted by $r_1,\ldots,r_{W*}$.
Moreover, let $p_i$ be the probability that $\po_k$ is identical to $r_i$.
Note that $W^*,p_i,r_i$ are all unknown and we do not build $\mathcal{A}$ explicitly.
Remind that the entropy $H(\po_k)$ is simply defined as $\sum_i p_i \log (1/p_i)$.

In learning phase, we take $T\geq W\log n$ instances to sample the results of $\po_k$ and count their frequency.
For $1\leq i\leq W^*$, denote by $\chi_i$ the times that $r_i$ is sampled. Let $q_i=\chi_i/T$.
(Note that $\chi_i$ might be zero for some $r_i$; such $r_i$ is unknown to us. Other $r_i$'s are known.)

\subsection{Store all the sampled results of $\po_k$ in a trie}

We encode every known result of $\po_k$ by a vector $(b_1,\ldots,b_{n_k})$ (similar to the Lehmer code).
\begin{definition}\label{def:encoding}
Given a known result of $\po_k$, element $b_1$ is defined as among $V_0,\ldots,V_n$ the predecessor of $x_1$;
and $b_2$ is defined as among $V_0,\ldots,V_n,x_1$ the predecessor of $x_2$;
so on and so forth; finally, $b_{n_k}$ is defined as the predecessor of $x_{n_k}$ among $V_0,\ldots,V_n,x_1,\ldots,x_{n_k-1}$ .
\end{definition}

\begin{figure}[h]
  \centering \includegraphics[width=.9\textwidth]{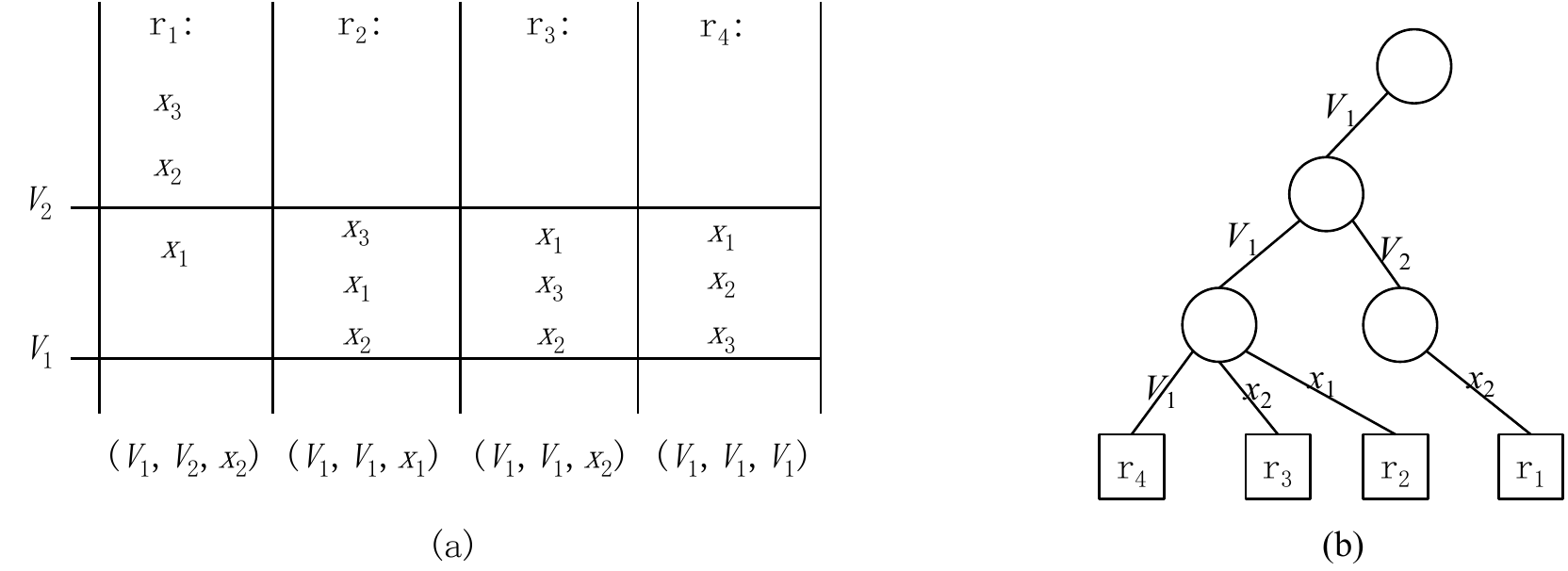}\\
  \caption{Illustration of the encoding given in Definition~\ref{def:encoding} and the trie.}\label{fig:trie}
\end{figure}

Four examples are given in Figure~\ref{fig:trie}~(a). The bottom of the columns shows the vectors.

We store the vectors of all sampled results of $\po_k$ into a trie as shown in Figure~\ref{fig:trie}~(b).
Moreover, we assign every node in this trie a weight: A leaf labeled by $r_i$ has weight $q_i$,
  and the weight of an internal node equals the total weight of its sons; so the root has weight 1.

\section{Operation phase Step 1 -- compute $\po_k$}

First, let us consider an ideal case where $q\equiv p$, i.e. $q_i=p_i$ for every $1\leq i\leq W^*$,.

Assume we are given the values of $(x_1,\ldots,x_{n_k})$ and we want to determine $\po_k$.
  Equivalently, we want to determine the vector corresponding to $\po_k$.
  Similar as what Fredman did in \cite{FREDMAN76},
    using $(x_1,\ldots,x_{n_k})$,
    we can compute $b_1,\ldots,b_{n_k}$ step by step. When $\po_k=r_i$, this process corresponds to a path in the trie starting from the root to the leaf labeled with $r_i$.

  According to some basic algorithmic knowledge (see section~3.2 paragraph 1 in \cite{best}),
    if currently we are at a node with weight $w_j$ and the next round we proceed to a son with weight $w_k$,
    the time for choosing the son in this step would be $O(1+\log(w_j/w_k))$.
  Therefore, if $\po_k=r_i$, it takes $O(n_k+\log(1/q_i))$ time to reach the node labeled with $r_i$.

  Further since the probability that ``$\po_k=r_i$'' is $p_i$,
  the expected time for computing $\po_k$ would be $O(\sum_i p_i(n_k+\log(1/q_i)))=O(n_k+\sum_i p_i\log (1/q_i))=O(n_k+H(\po_k))$ when $q\equiv p$.

\smallskip Next, we show that even if $q\neq p$, the expected running time is still $O(n_k+H(\po_k))$.

\subsection{The proof of Lemma~\ref{lemma:H_k}}

\newcommand{\bq}{\mathbf{q}}

Denote $\bq=(q_1,\ldots,q_{W^*})$.
Let $t^\bq_i$ be the time for computing $\po_k$ when $\po_k=r_i$ and when our sampling result is some fixed $\bq$.
Similar as in the above case, for $q_i>0$, we compute $\po_k$ in time $O(n_k+\log(1/q_i))$ when $\po_k=r_i$;
yet for $q_i=0$, we find no result after searching the trie and we use a trivial method to compute $\po_k$ and it costs $O(n_k\cdot\log n)$ time.
Therefore,
\begin{equation}
t^\bq_i=\left\{
               \begin{array}{ll}
                 O(n_k+\log(1/q_i)), & q_i>0; \\
                 O(n_k\cdot \log n), & q_i=0.
               \end{array}
             \right.
\end{equation}

Thus the expected running time for computing $\po_k$ in operation phase is given by
\begin{equation}
\begin{aligned}
& \sum_\bq \Pr(\bq) \cdot \sum_i p_i t^\bq_i = \sum_i p_i \sum_\bq \Pr(\bq) t^\bq_i\\
& =\sum_i p_i \sum_{\bq:q_i>0} \Pr(\bq) O\big(n_k+\log(1/q_i)\big) + \sum_i p_i \sum_{\bq:q_i=0} \Pr(\bq) O \big( n_k \log n \big)
\end{aligned}
\end{equation}
\begin{equation}
\hbox{The second term}=O\big(n_k\log n \sum _i p_i (1-p_i)^T\big)\leq O\big(n_k\log n W^*/(T+1)\big)=O(n_k).
\end{equation}
\begin{equation}
\begin{aligned}
\hbox{The first term}& = \sum_i p_i \sum_{\bq:q_i>0} \Pr(\bq) O(n_k)  + \sum_i p_i \sum_{\bq:q_i>0} \Pr(\bq) O\big(\log(1/q_i)\big)\\
                     & \leq O(n_k)  + \sum_i p_i \sum_{j=1}^{T}\Pr(q_i=\frac{j}{T}) O\big(\log(\frac{T}{j})\big)
\end{aligned}
\end{equation}
\begin{equation}
\begin{gathered}
 \sum_i p_i \sum_{j=1}^{T}\Pr(q_i=\frac{j}{T}) O\big(\log(\frac{T}{j})\big)\\
=\sum_i p_i \sum_{1\leq j\leq p_iT/2}\Pr(q_i=\frac{j}{T}) O\big(\log(\frac{T}{j})\big)
+\sum_i p_i \sum_{p_iT/2<j\leq T}\Pr(q_i=\frac{j}{T}) O\big(\log(\frac{T}{j})\big)\\
\leq \sum_i p_i \sum_{1\leq j\leq p_iT/2}\Pr(q_i=\frac{j}{T}) O\big(\log T\big)
+\sum_i p_i \sum_{p_iT/2<j\leq T}\Pr(q_i=\frac{j}{T}) O\big(\log(\frac{2}{p_i})\big)
\end{gathered}
\end{equation}
\begin{equation}
\hbox{The second term}\leq \sum_i p_i O\big(\log(2/p_i)\big)=O(1+H(\po_k)).
\end{equation}

To bound the first term, we need to bound $\sum_{1\leq j\leq p_iT/2}\Pr(q_i=\frac{j}{T})<\Pr(q_i\leq p_i/2)$, for which
we apply the Chernoff bound. Note that the expectation of $q_i$ is given by $p_i$, so $\Pr(q_i\leq p_i/2)\leq e^{-p_iT/8}\leq \frac{8}{p_iT}$.
Hence $\hbox{the first term}\leq \sum_i p_i \frac{8}{p_iT} O(\log T)=O(W'\log T/T)=O(1)$.

To sum up, altogether we prove that the expected running time is $O(n_k+H(\po_k))$.

\bibliographystyle{plainurl}
\bibliography{self}

\end{document}